\documentclass{aa}  

\usepackage{graphicx}
\usepackage{txfonts}
\usepackage[colorlinks=true, allcolors=blue]{hyperref}
\begin{document}

   \title{\texttt{sunset}: A database of synthetic atmospheric-escape transmission spectra for nearly every transiting planet}

   \author{Dion Linssen \thanks{E-mail: d.c.linssen@uva.nl}
          \inst{1}
          \and
          Antonija Oklop\v{c}i\'{c}\inst{1}
          \and
          Morgan MacLeod\inst{2}
          }

   \institute{Anton Pannekoek Institute for Astronomy, University of Amsterdam,
              Science Park 904, 1098 XH Amsterdam, The Netherlands
         \and
             Center for Astrophysics | Harvard \& Smithsonian, 60 Garden Street, MS-16, Cambridge, MA 02138, USA
             }

   \date{Received; accepted}

  \abstract
   {Studying atmospheric escape from exoplanets can provide important clues about the formation and evolution of exoplanets. Observational evidence of atmospheric escape has been obtained through transit spectroscopy in strong spectral lines of various atomic species. In recent years, the number of exoplanets that have been targeted in this way has grown rapidly, mainly by observations of the metastable helium triplet. Even with this larger sample of exoplanets, many aspects of atmospheric escape remain not fully understood, such as the role of the stellar high-energy spectrum and planetary magnetic field, highlighting the need for additional observations. This work aims to identify the best targets for observations in various spectral lines. Using the atmospheric escape code \texttt{sunbather}, we calculate a synthetic transmission spectrum of nearly every transiting exoplanet currently known. This database of spectra, named \texttt{sunset}, is publicly available. We introduce metrics based on the spectral line strengths and system distance or magnitude, which allow swift identification of the most favorable targets. By analyzing the complete set of spectra from a demographic perspective, we find that the strengths of many spectral lines do not correlate strongly with the atmospheric mass-loss rate, suggesting that a nondetection does not immediately rule out an escaping atmosphere. Some lines also do not correlate very strongly with one another, emphasizing the benefits of observing multiple spectral lines. Our model spectra show only a weak correlation between the XUV (X-ray and extreme UV) flux and the helium line strength, affirming that the absence of such a trend found by observational works is in fact as expected. A direct comparison between our synthetic spectra and the sample of observed metastable helium spectra shows that they are generally consistent within the large model uncertainties. This suggests that by and large, photoevaporation is able to explain the current metastable helium census.}

   \maketitle
%

\section{Introduction} \label{sec:introduction}
Thousands of exoplanets have been discovered to date. The majority of them have been discovered with the transit and radial velocity methods, which are biased towards finding planets at short orbital distances. Therefore, most of the currently known exoplanets receive high amounts of stellar radiation, which can cause their atmospheres to escape \citep[e.g.,][]{yelle_aeronomy_2004, koskinen_stability_2007}. For some planets, the escape process can be so efficient that it strips a significant part of the planet's mass. This is theorized to be one of the mechanisms responsible for shaping the sub-Jovian desert and radius valley \citep{szabo_short-period_2011, kurokawa_mass-loss_2014, fulton_california-_2017, owen_photoevaporation_2018, ho_deep_2023}. High-gravity and further-out planets are believed to be more stable against transformative mass-loss. In addition to the evolutionary implications, studying atmospheric escape can also provide valuable insights into physical phenomena such as planetary magnetic fields and stellar winds \citep{bourrier_3d_2013, vidotto_exoplanets_2017, carolan_effects_2021, gupta_impact_2023}, which are known to affect atmospheric mass loss in solar system planets \citep[e.g.,][]{gronoff_atmospheric_2020}.

Ongoing atmospheric escape can be studied with transit spectroscopy in spectral lines that form in the low-density, high altitude layers of the atmosphere, such as the hydrogen Lyman-$\alpha$ line \citep[e.g.,][]{vidal-madjar_extended_2003, lecavelier_des_etangs_evaporation_2010, bourrier_hubble_2018}, the metastable helium triplet \citep[e.g.,][]{oklopcic_new_2018, spake_helium_2018, allart_spectrally_2018, nortmann_ground-based_2018}, and various lines from metals, mostly in the UV \citep[e.g.,][]{vidal-madjar_detection_2004, sing_hubble_2019, garcia_munoz_heavy_2021}. In recent years, the number of planets which have been targeted in the metastable helium triplet has grown rapidly. It has thus started to become feasible to interpret them in a statistical sense, in addition to the more in-depth interpretation and modeling of individual planets. Several works have already searched for correlations between the observed helium signals and system parameters that should influence atmospheric escape, such as the age, high-energy radiation, stellar spectral type, and planetary gravity \citep{zhang_outflowing_2023, bennett_nondetection_2023, krishnamurthy_helium_2024, orell-miquel_mopys_2024}. The expected preference of helium detections around K-type host stars has in this way been observationally supported \citep{oklopcic_helium_2019, krishnamurthy_helium_2024, orell-miquel_mopys_2024}. However, the expected dependence on the X-ray and extreme UV (EUV; together XUV) flux is somewhat unclear, as \citet{bennett_nondetection_2023} found a correlation with the stellar $R'_{HK}$ activity indicator, while no correlation with the stellar age has been found \citep{krishnamurthy_helium_2024, orell-miquel_mopys_2024}, although both parameters are related to the stellar XUV flux.

To better determine the role that each of these system parameters play, a larger sample of atmospheric escape observations is needed. In this regard, the upcoming high-resolution spectrograph NIGHT will prove extremely useful, as it will be dedicated to metastable helium triplet observations of more than a hundred exoplanets \citep{farretjentink_night_2024, farret_jentink_near-infrared_2024}. At the same time, it will be helpful to increase the sample of exoplanets with observations in other spectral lines that trace atmospheric escape. Other spectral lines will have a different dependence on the same system parameters \citep[e.g.,][]{linssen_expanding_2023}, so that by observing these lines we may get a more comprehensive view of the physics that govern atmospheric escape, rather than the physics underlying the metastable helium triplet specifically.

One of the main goals of this work is to identify exoplanets that are potentially good targets in upcoming observations of various spectral lines that trace atmospheric escape. Our approach is based on using \texttt{sunbather} \citep{linssen_open-source_2024}, an open-source code for modeling the spectral signatures of escaping atmospheres. We calculate a synthetic transmission spectrum of the upper atmosphere of almost every transiting exoplanet currently known. The synthetic spectra are freely available to the community in our online database called \texttt{sunset}\footnote{The \texttt{sunset} database: \url{https://zenodo.org/records/13785440} \label{fn:data}}, and can be used to inform observational campaigns. 

This paper is structured as follows. In Sect. \ref{sec:methods}, we describe our numerical modeling, including the input system parameters, stellar SED templates, and free model parameters. We also provide an overview of the assumptions and sources of uncertainty in our models. In Sect. \ref{sec:results}, we describe our resulting output spectra, and explore correlations between various system parameters and spectral line strengths. In Sect. \ref{sec:comparison_to_sample}, we compare our synthetic spectra to the sample of planets with metastable helium observations. In Sect. \ref{sec:summary}, we summarize. Additionally, in Appendix \ref{app:good_targets}, we list the most promising observational targets for a selection of spectral lines.

\section{Methods} \label{sec:methods}
\subsection{System parameters}
We start by downloading the planetary parameters from the NASA Exoplanet Archive\footnote{\url{https://exoplanetarchive.ipac.caltech.edu}}. We use the Composite Planet Data from January 16, 2024. The composite data combines multiple publications into one set of parameters per planet, meaning that there is relatively high completeness of the parameters, but not all of them are internally consistent. From the initial database of 5569 planets, we first select the 4182 planets that are transiting and noncontroversial, and then select the 4144 planets (hereafter, our ``initial population'') that have a reported value for all parameters that we need to produce model spectra. Required are the planet mass ($M_p$), radius ($R_p$), and semi-major axis ($a$), as well as the stellar mass ($M_*$), radius ($R_*$), and a value for either the star's spectral type or effective temperature ($T_{\rm eff}$, which we then use as a proxy for the spectral type, assuming the star is on the main sequence). For many planets, some of these parameters have considerable uncertainties, which propagate into our model results. Instead of filtering out such planets beforehand, we proceed to model every planet possible, and provide tailored warnings for individual planets that have large uncertainties on used parameters. 

\subsection{Stellar SED templates}
\begin{figure*}
   \centering
   \includegraphics[width=\hsize]{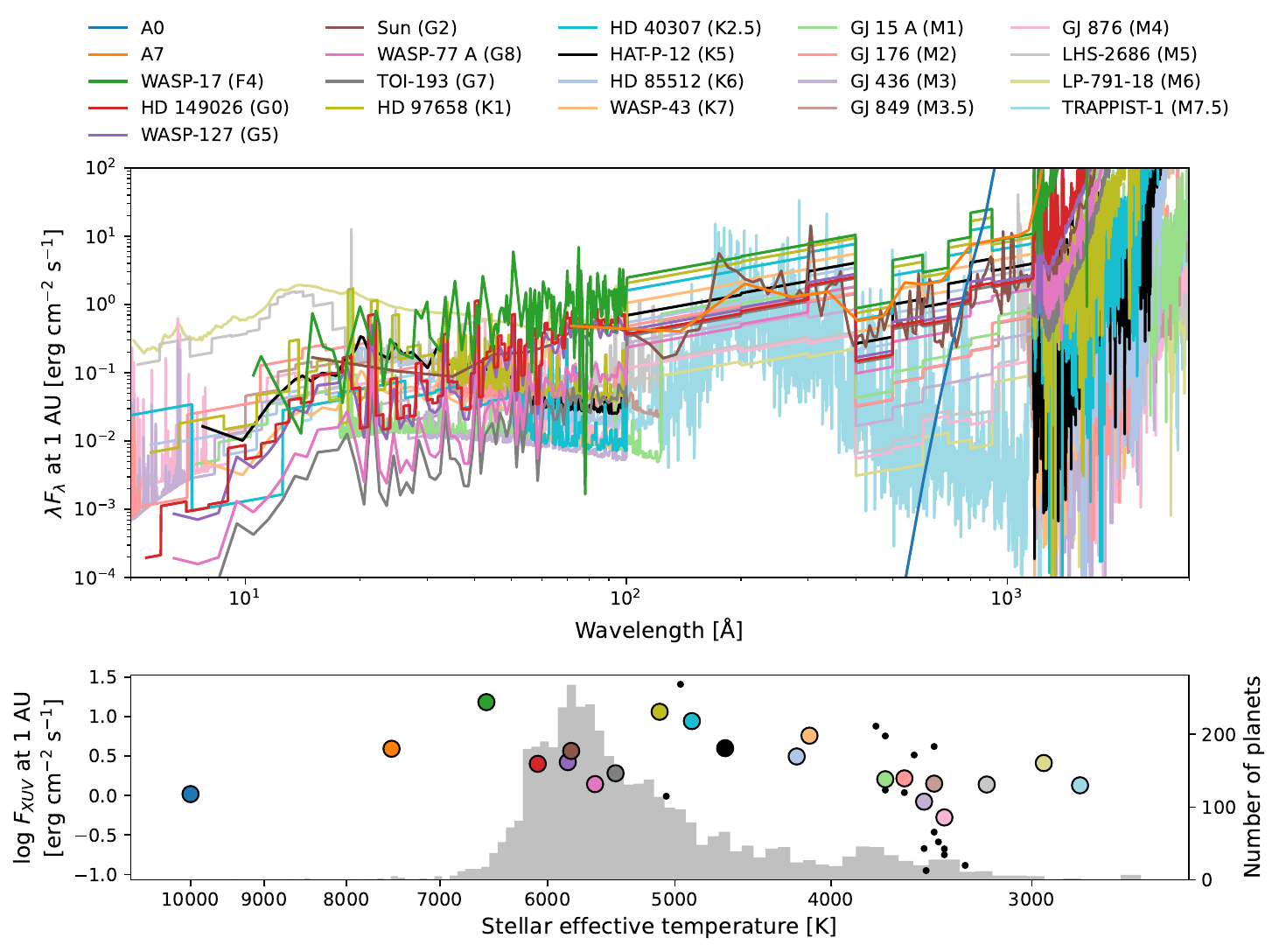}
      \caption{{\it Top:} Stellar spectral energy distribution (SED) templates used in this work. The A0 and A7 spectra are based on the work of \citet{fossati_extreme-ultraviolet_2018}, the solar spectrum is from the TIMED and SORCE missions \citep{woods_solar_2005, rottman_sorce_2005}, and the rest of the spectra are from the MUSCLES survey \citep{france_muscles_2016, youngblood_muscles_2016, loyd_muscles_2016, wilson_mega-muscles_2021, behr_muscles_2023}. {\it Bottom:} The integrated XUV flux ($\lambda < 911$~Å) of each SED, as a function of the effective temperature of the star (scatter points; see also Table \ref{tab:stellar_SEDs}). Small black points indicate SEDs in the MUSCLES database that we do not use. The grey histogram (read from the right y-axis) shows the host star effective temperatures of the initial planet population. Planets are assigned the SED template of the closest stellar spectral type. If the spectral type is not reported, the SED template of the star with the closest effective temperature is assigned.}
        \label{fig:stellar_SEDs}
   \end{figure*}

One of the most important yet uncertain factors affecting the planet's mass-loss rate, upper atmospheric structure, and transit spectrum, is the host star's spectral energy distribution (SED), particularly at XUV wavelengths. These SEDs are not readily available for the great majority of exoplanets. Therefore, we use the SEDs of proxy stars with a similar spectral type. Our sample of stellar SEDs mainly consists of spectra from the MUSCLES, Mega-MUSCLES and MUSCLES Extension surveys \citep{france_muscles_2016, youngblood_muscles_2016, loyd_muscles_2016, wilson_mega-muscles_2021, behr_muscles_2023}. This is a collection of panchromatic spectra of stars ranging from F- to M-type, with measured X-ray and reconstructed EUV components. For some spectral types, mainly between M1 and M4, there are multiple stars in the survey with a similar spectral type, but varying levels of XUV flux ($F_{XUV}$). For planets in our initial population with such host spectral type, it is generally not clear which SED template is most applicable to use. Therefore, starting from the full collection of 34 available MUSCLES spectra, we selected a subset of 18 stars with non-overlapping spectral types. We selected stars with intermediate levels of XUV flux, thereby aiming to exclude relatively active or inactive stars (see lower panel of Fig. \ref{fig:stellar_SEDs}). We then supplemented this sample with the SED of our Sun, based on measurements by the TIMED \citep{woods_solar_2005} and SORCE \citep{rottman_sorce_2005} spacecrafts, due to the high data quality of this spectrum. The way the solar spectrum was constructed is described in more detail in \citet{linssen_open-source_2024}. Finally, to accommodate earlier-type stars, we added the spectra of an A7 and an A0 star. These are based on the model spectra with effective temperatures of 7500~K and 10\;000~K, respectively, from \citet{fossati_extreme-ultraviolet_2018}. The XUV flux of the A7 star is from a solar spectrum scaled up by a factor of three. The full sample of used SEDs with their spectral types and effective temperatures are listed in Table \ref{tab:stellar_SEDs} and shown in Fig. \ref{fig:stellar_SEDs}. 

In our initial planet population, there are 944 planets with a reported host spectral type, and we assign the SED with the closest spectral type to them. The remaining planets all have a reported stellar effective temperature, and we assign the SED with the closest effective temperature to them. This is more or less the simplest approach possible, which does not factor in the age and activity levels of the planet host star, or more generally, the possible variety in XUV flux level of the star.

\begin{table}
\caption{Spectral types and effective temperatures of the stars used in our SED sample. For the stars from the MUSCLES Extension \citep{behr_muscles_2023}, we adopted the spectral types and effective temperatures as listed in the paper. For the other stars, we first adopted the spectral type from Simbad\tablefootmark{1}, using an intermediate value when there are multiple measurements. We then adopted an effective temperature based on a conversion table from the spectral type, assuming main-sequence stars\tablefootmark{2}.}
\label{tab:stellar_SEDs}
\centering
\begin{tabular}{l c c}
\hline\hline
Star & Spectral type & Effective temperature [K]\\
\hline                    
   -\tablefootmark{3} & A0 & 10\;000\\
   -\tablefootmark{4} & A7 & 7500\\
   WASP-17 & F4 & 6548\\
   HD 149026 & G0 & 6084\\
   WASP-127 & G5 & 5828\\
   Sun & G2 & 5800\\
   WASP-77 A & G8 & 5605\\
   TOI-193 & G7 & 5443\\
   HD 97658 & K1 & 5110\\
   HD 40307 & K2.5 & 4880\\
   HAT-P-12 & K5 & 4653\\
   HD 85512 & K6 & 4200\\
   WASP-43 & K7 & 4124\\
   GJ 15 A & M1 & 3700\\
   GJ 176 & M2 & 3600\\
   GJ 436 & M3 & 3500\\
   GJ 849 & M3.5 & 3450\\
   GJ 876 & M4 & 3400\\
   LHS-2686 & M5 & 3200\\
   LP-791-18 & M6 & 2949\\
   TRAPPIST-1 & M7.5 & 2800\\
\hline
\end{tabular}
\tablefoot{
\tablefoottext{1}{\url{https://simbad.u-strasbg.fr/simbad/}}
\tablefoottext{2}{\url{https://sites.uni.edu/morgans/astro/course/Notes/section2/spectraltemps.html}}
\tablefoottext{3}{Synthetic spectrum from \citet{fossati_extreme-ultraviolet_2018}.}
\tablefoottext{4}{Synthetic spectrum from \citet{fossati_extreme-ultraviolet_2018}, with 3x solar XUV flux.}
}
\end{table}

\subsection{Model parameters} \label{sec:model_parameters}
Our simulations are performed using the 1D atmospheric escape code \texttt{sunbather} \citep{linssen_open-source_2024}. This code first models the upper atmosphere as an isothermal Parker wind \citep{parker_dynamics_1958, lamers_introduction_1999} using the \texttt{p-winds} \citep{dos_santos_p-winds_2022} and \texttt{Cloudy} \citep{ferland_2017_2017, chatzikos_2023_2023} codes. It afterwards refines the temperature to a nonisothermal profile. The use of \texttt{Cloudy} then enables the calculation of transmission spectra spanning a large wavelength range and including spectral lines from many different elements. 

\begin{figure}
   \centering
   \includegraphics[width=\hsize]{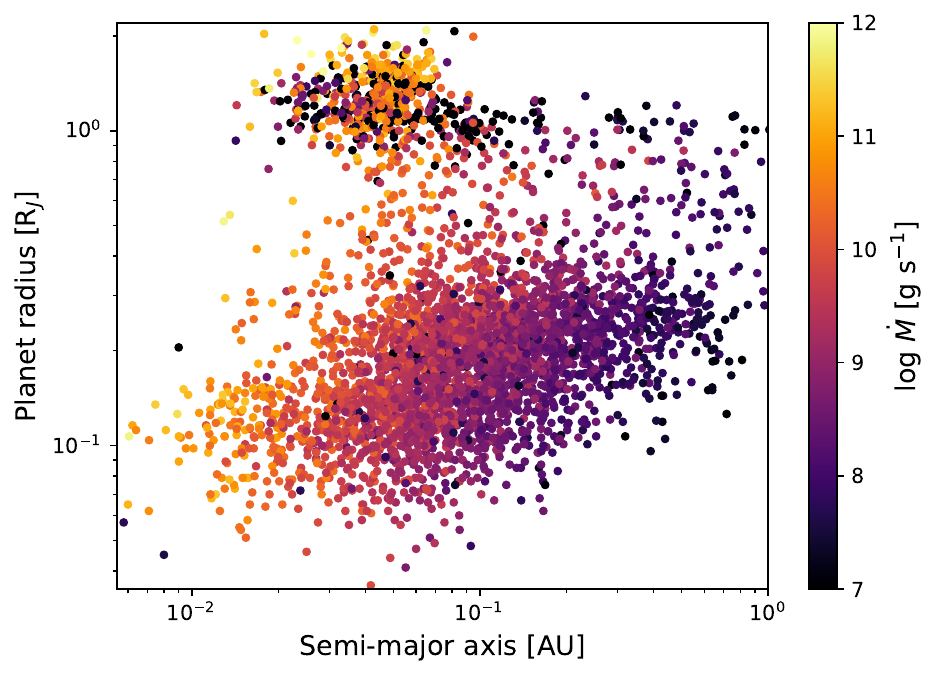}
      \caption{Mass-loss rates of the initial planet population, calculated from the analytical approximation of \citet[][Appendix A]{caldiroli_irradiation-driven_2022}. The mass-loss rate is parametrized as a function of $R_p$, $M_p$, $a$, $M_*$, and $F_{XUV}$, based on the output of the first-principle 1D photoevaporation code \texttt{ATES} \citep{caldiroli_irradiation-driven_2021}.}
         \label{fig:mdot_population}
   \end{figure}

In addition to the system parameters and the SED, three main free parameters are required to run \texttt{sunbather}: the mass-loss rate ($\dot{M}$), the temperature ($T_0$), and the atmospheric composition. To choose the mass-loss rate of each planet, we use the parametrization given in Appendix A of \citet{caldiroli_irradiation-driven_2022}. They ran a grid of models with the \texttt{ATES} code \citep{caldiroli_irradiation-driven_2021}, which simulates the launching of photoevaporative outflows by modeling photoionization of hydrogen-helium atmospheres and predicts the corresponding mass-loss rates. Their analytic formula for the mass-loss rate is calculated for a hydrogen-helium atmosphere (with relative abundances set to He/H=0.083 in number density), expressed as the energy-limit escape formula, where the parametrization takes place in the form of a non-constant efficiency factor $\eta_{\rm eff}$. However, we clarify that these mass-loss rates should not be thought of as being energy-limited. \texttt{ATES} does not strictly model outflows that are in the energy-limited regime, and is perfectly able to transition into a recombination-limited outflow, for example. In such a case, the efficiency factor would simply become much lower than one. Their parametrization of the mass-loss rate comes from fitting a complex functional form to the output mass-loss rates of the model grid, with dependencies on $R_p$, $M_p$, $a$, $M_*$, and $F_{XUV}$. This analytic fit is precise to within a factor $\sim$1.5, but can only safely be assumed to be valid in their explored parameter range: $10^2 \lesssim F_{XUV} / \rho_p \lesssim 10^6$ and $10^{12.17} \lesssim K \phi \lesssim 10^{13.29}$, where $\rho_p$ is the planet's mean density, $K$ is the Roche-lobe correction factor from \citet{erkaev_roche_2007}, and $\phi=GM_p/R_p$ is the planet's gravitational potential. Unfortunately, there are quite many planets in our initial population that fall outside of these bounds. Most of these are cases where $K \phi$ is below its lower bound (and often $F_{XUV}/\rho_p$ too), which is a part of the parameter space where the efficiency varies only slowly with both quantities (see Fig. A.1 of \citealt{caldiroli_irradiation-driven_2022}), so that we are still relatively confident in applying the parametrization outside of these bounds. There is also a smaller number of planets where $K \phi$ is above its upper bound, where the efficiency dependency is strong. Our adopted efficiency (and hence mass-loss rate) for these planets will be much more inaccurate. These high-gravity planets represent the non-inflated Jovian population. However, the efficiency already tends to $\eta_{\rm eff} \lesssim 10^{-4}$ at this upper boundary, leading to very low mass-loss rates. At low mass-loss rates, \texttt{sunbather} often fails to converge (the reasons for this will be discussed in more detail below), so that most of these planets do not make it to our successfully simulated ``final population'' anyway. In Fig. \ref{fig:mdot_population}, we show the mass-loss rates we obtain for the planets in our initial population.

The second free parameter of \texttt{sunbather} is the temperature. To determine which value to use, we do a ``self-consistency check'', where we compare the chosen $T_0$ to the nonisothermal temperature profile that the code solves for. Such a self-consistency check has been performed before, albeit using different criteria \citep{linssen_constraining_2022, linssen_expanding_2023, linssen_open-source_2024}. Here, we search for the value of $T_0$ that is equal to the maximum temperature of its nonisothermal profile ($\hat{T}$). Such a value of $T_0$ is unique, meaning there is only one possible value (at a given mass-loss rate). Multiple simulations with different values of $T_0$ are typically needed to find this ``self-consistent'' $T_0$. In concrete terms, we start the search for $T_0$ of each planet with a simulation at $T_0=10,000$~K. We then extract $\hat{T}$ from the nonisothermal profile and start a new simulation at $T_0$ equal to $(\hat{T}+T_0)/2$. We are satisfied when the temperatures are self-consistent to within $|\hat{T} - T_0| < 300$~K. Afterwards, we only use the ``self-consistent'' model and discard the other ones, so that we have one atmospheric model per planet.

Since \texttt{sunbather} involves the use of \texttt{Cloudy}, it is subject to the limitations of \texttt{Cloudy}. At high densities (roughly $\gtrsim 10^{15}$~g~s$^{-1}$), \texttt{Cloudy} will often fail to converge, as the code was not designed for such conditions. In general, such high densities are reached for high-gravity planets, and/or at low $T_0$. The latter is the reason that we use a self-consistency criterion based on the maximum of the nonisothermal temperature profile, since that yields the highest values of $T_0$ for which one may reasonably argue that the temperature is self-consistent. In this way, we are able to successfully simulate more planets, but it does admittedly introduce some bias towards lower density, higher temperature and velocity outflows. Using this criterion, the highest gravity planets still reach atmospheric densities above what \texttt{Cloudy} can handle, and we thus fail to simulate them. Fortunately, these high-gravity planets are also expected to be stable against atmospheric escape, and therefore not very interesting targets in the context of our research.

The final free model parameter is the atmospheric composition. We use a solar composition for all planets. We motivate this by the fact that the mass-loss rate predictions from \texttt{ATES} are based on hydrogen-helium only simulations, and would not be valid for high-metallicity outflows. We still add metals at solar abundances to our simulations, because they do not affect the atmospheric structure very significantly at these relatively low abundances \citep[e.g.,][]{zhang_escaping_2022, kubyshkina_precise_2024, linssen_open-source_2024}, but already produce strong lines in the transit spectrum that we aim to investigate. A high metallicity would likely have resulted in a different atmospheric structure, mass-loss rate, and transit spectrum, but running models with different compositions for each planet is beyond the scope of this work. 

\subsection{Calculation of transmission spectra}
With mass-loss rates calculated and an algorithm in place to determine the temperature, we start the \texttt{sunbather} simulations for all 4144 planets. 220 of them fail because of high-density problems with \texttt{Cloudy}, while another 26 fail because of instabilities in the \texttt{sunbather} algorithm that solves the nonisothermal temperature profile. We successfully simulate the remaining 3898 planets (hereafter, our ``final population''), and continue to calculate their transmission spectra. We use the planet's own impact parameter, or 0 if it is not reported. The spectra are calculated at mid-transit, with no stellar center-to-limb variations. The wavelength grid runs from 911~Å to 11,000~Å at a high spectral resolution of $R\sim400,000$. A total of 12,191 spectral lines from various elements are present in this wavelength region. All gas within the atmospheric domain of [$1R_p$, $8R_p$] is included (the part of it that covers the stellar disk), irrespective of the Roche radius. This choice is somewhat arbitrary and can be debated. Beyond the Roche radius, our 1D model most certainly does not capture the geometry of the escaping material. However, since escaping gas does not simply disappear at the Roche radius, we opt to include gas beyond it, to allow for the existence of material in tails and streams contributing to the line formation. Lines that generally form below the Roche radius are not very affected by this choice, while lines that form at high altitudes are more uncertain (see \citealt{linssen_expanding_2023} for an overview of the strongest lines and at which altitudes they form).

\subsection{Uncertainties and warnings} \label{sec:warnings}
Having described our modeling approach, we here provide a concise overview of the numerous uncertainties in our methods. Some of these uncertainties apply to most or all planets, and they have to do with choices and assumptions in our modeling framework. These overall uncertainties include:
\begin{itemize}
    \item The planet parameters are often collected from multiple sources, which are potentially not mutually consistent.
    \item The XUV flux received by practically every planet is highly uncertain. We use template spectra from proxy host stars. These spectra may not be representative of the true flux received by the planet.
    \item The planet is assumed to have an atmosphere.
    \item The atmospheric composition may be different from solar.
    \item The upper atmosphere may not be well described by a 1D Parker wind with a mass-loss rate as predicted by \texttt{ATES}, for example when atmospheric escape is not driven by photoevaporation, or when there are significant day-to-nightside differences, magnetic fields, stellar wind interactions, etc.
    \item The chosen $T_0$-value may be inappropriate. The criterion to choose its value by requiring self-consistency with the maximum temperature in the nonisothermal profile is somewhat arbitrary.
    \item The transmission spectra are made without stellar center-to-limb darkening. Limb-darkening (or limb-brightening at FUV wavelengths, see \citealt{schlawin_exoplanetary_2010}) could have a minor effect.
\end{itemize}

In addition to these methodological uncertainties, there are additional uncertainties or limiting factors that only apply to specific planets, typically because one or more of the input parameters are not well constrained. These include (the quoted planet numbers only include the successfully simulated planets of our final population):
\begin{itemize}
    \item Molecules were present at non-negligible abundances in the \texttt{Cloudy} simulation, which affects the atmospheric structure but is ignored by \texttt{sunbather} (7 planets).
    \item The planet has eccentricity larger than 0.1 (296 planets), or the planet does not have a reported eccentricity (455 planets). \texttt{sunbather} assumes a circular orbit.
    \item The planet radius errors are larger than 10\% (1723 planets), or the errors are not reported (431 planets).
    \item The planet mass quoted in the NASA Exoplanet Archive is derived from the radius through a mass-radius relationship (2845 planets).
    \item The planet mass is actually $M_p {\rm sin}(i)$ or $M_p {\rm sin}(i) / {\rm sin}(i)$ (4 planets).
    \item The planet mass errors are larger than 20\% (359 planets), or the errors are not reported (81 planets).
    \item We calculated the semi-major axis from a reported $a/R_*$ value (221 planets).
    \item The semi-major axis errors are larger than 25\% (39 planets), or the errors are not reported (2428 planets).
    \item The stellar radius errors are larger than 25\% (612 planets), or the errors are not reported (72 planets).
    \item The stellar mass errors are larger than 25\% (129 planets), or the errors are not reported (81 planets).
    \item The planet transit impact parameter is unknown and we assumed 0 (135 planets).
    \item The host star's spectral type is more than two away from that of the used SED (e.g., F1 versus F4; 19 planets).
    \item The host star's spectral type is not reported and the effective temperature was used instead (3069 planets).
    \item The host star's effective temperature is more than 200~K away from that of the used SED (83 planets).
    \item The stellar age is unknown. The star is potentially young with higher XUV flux (634 planets).
    \item The star is younger than 1~Gyr, with potentially higher XUV flux (362 planets).
    \item The $F_{XUV}/\rho_p$ value is outside of the formally valid bounds of the mass-loss rate parametrization (1490 planets).
    \item The $K \phi$ value is outside of the formally valid bounds of the mass-loss rate parametrization (1963 planets).
    \item The planet's bulk density is higher than 3~g~s$^{-1}$, which may indicate that it is a rocky planet, which likely does not have a primordial extended atmosphere (1769 planets).
\end{itemize}

Out of these different warnings, we consider the following to be relatively important: the mass is from a mass-radius relationship, no or large errors on the planet radius, planet mass, semi-major axis, or stellar mass, a used SED template that is more than two spectral types or 200~K effective temperature away from the host star, calculation of the mass-loss rate outside the validity bounds of the \citet{caldiroli_irradiation-driven_2022} formula, and planets with bulk density $>3$~g~s$^{-1}$ that are potentially rocky. There are 326 exoplanets in our final population that have none of these warnings, and their model spectra are expected to be most accurate.

\section{Results} \label{sec:results}
\begin{figure*}
   \centering
   \includegraphics[width=\hsize]{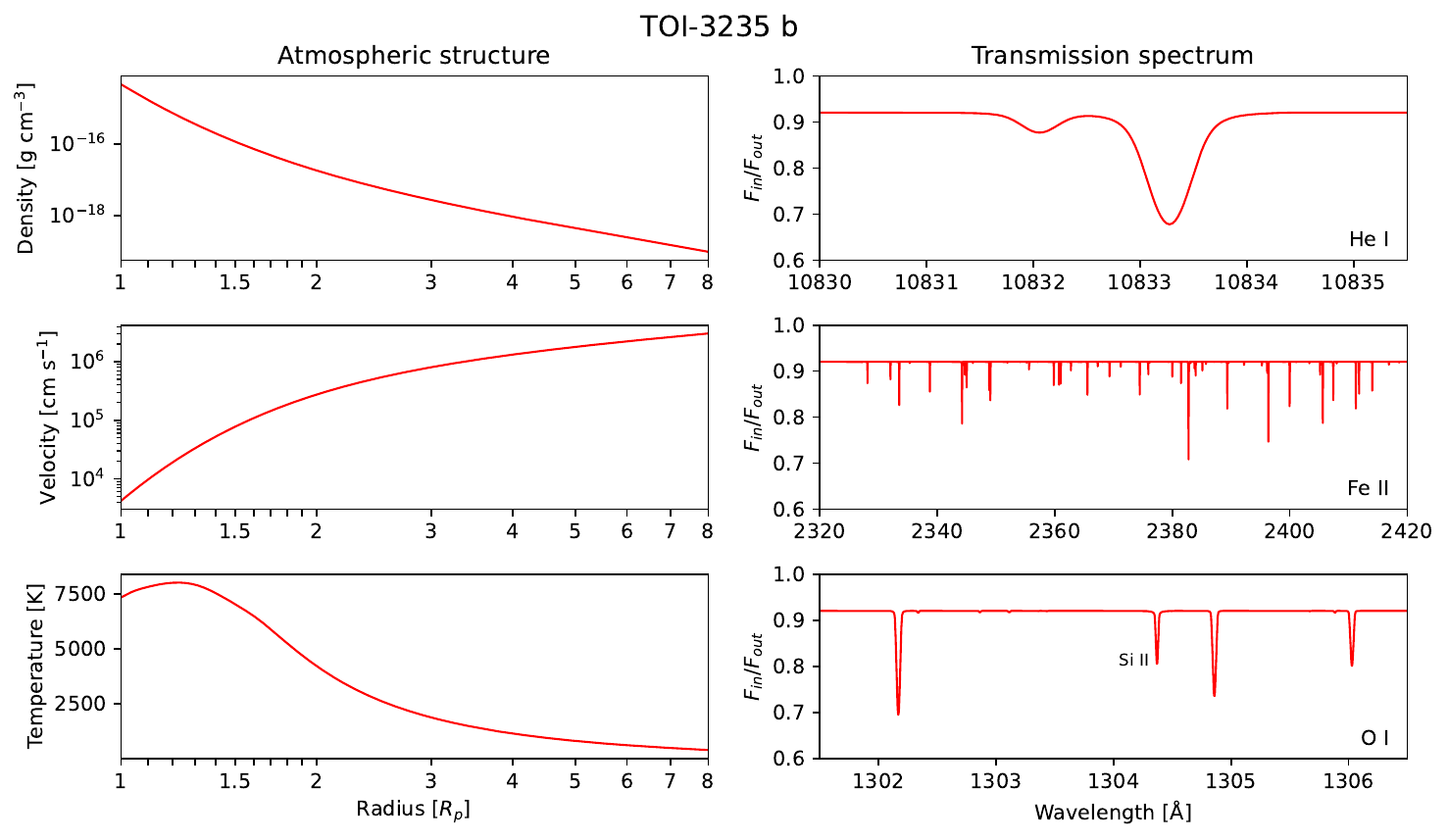}
      \caption{Example of the publicly available$^{\ref{fn:data}}$ output data products, for the planet TOI-3235~b (simulated with \texttt{sunbather}; \citealt{linssen_open-source_2024}). {\it Left column:} Upper atmospheric structure. {\it Right column:} Zoom-in on various parts of the transmission spectrum. The full spectrum spans 911~Å to 11,000~Å at a spectral resolution of $R\sim400,000$. It includes spectral lines from the 30 lightest elements (up until zinc).}
         \label{fig:data_products_example}
   \end{figure*}

Our output consists of atmospheric models (radial profiles of density, velocity, temperature, and mean particle mass) and transmission spectra of the final population of exoplanets. These are freely available for download in the \texttt{sunset} database$^{\ref{fn:data}}$. An example of the data products for a single exoplanet is shown in Fig. \ref{fig:data_products_example}. Additionally, for each planet, we provide an information file with the specific warnings from the list in Sect. \ref{sec:warnings} that apply to it. For many planets, there are numerous serious warnings, and we very strongly caution against the blind use of such spectra. One could point out that it may not have been useful to proceed to simulate these planets. However, we prefer to obtain a final population as large as possible, so that if needed, highly uncertain models can always be filtered out afterwards by inspecting their specific warnings. Even though some of these spectra may indeed not prove realistic and useful for the real-world planets they were modeled after, they are still useful for theoretically investigating the behaviour of atmospheric escape properties and transit spectra, given these system parameters. It is exactly this sort of analysis that we focus on in this section.

\subsection{Bulk outflow parameter correlations} \label{sec:bulk_correlations}
\begin{figure*}
   \centering
   \includegraphics[width=\hsize]{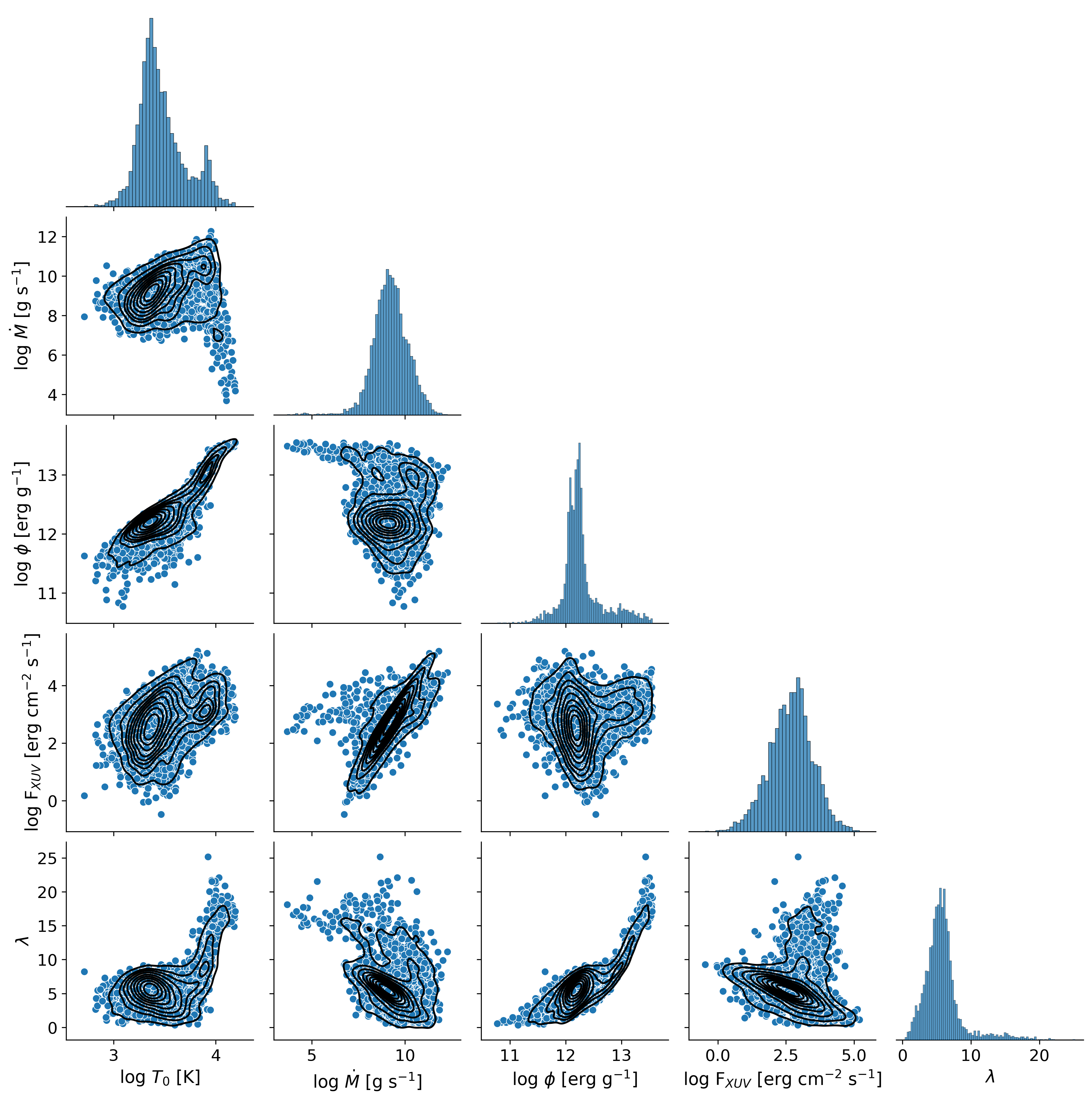}
      \caption{Correlations between various bulk parameters of the final planet population. The gravitational potential $\phi$ is calculated directly from the planet mass and radius as reported in the NASA Exoplanet Archive. The XUV flux $F_{XUV}$ comes from the stellar SED template. The mass-loss rate $\dot{M}$ is given by the \citet{caldiroli_irradiation-driven_2022} analytical approximation. The isothermal Parker wind temperature $T_0$ is a free model parameter whose value is determined through an iterative process. It satisfies the self-consistency criterion that it is equal to the maximum of the corresponding nonisothermal temperature profile (see Sec. \ref{sec:model_parameters}). The escape parameter $\lambda$ is calculated from $\phi$ and the sound speed, which depends on $T_0$ and the averaged mean particle mass $\bar{\mu}$ \citep[see][]{lampon_modelling_2020}.}
         \label{fig:correlations_bulkquantities}
   \end{figure*}

We start by investigating correlations between the various in- and output parameters, in order to gain insight into the physics controlling atmospheric escape. In Fig. \ref{fig:correlations_bulkquantities}, we show the correlations between five bulk quantities of the final planet population: the atmospheric temperature, the mass-loss rate, the planet gravitational potential, the XUV flux, and the escape parameter
\begin{equation}
    \lambda = \frac{\phi}{v_s^2} = \frac{G M_p}{R_p} \frac{\bar{\mu}}{k T_0},    
\end{equation}
where $v_s$ is the speed of sound, which is constant in the isothermal Parker wind model, and $\bar{\mu}$ is the averaged mean particle mass of the atmosphere \citep{lampon_modelling_2020}. 

When looking at the correlations in Fig. \ref{fig:correlations_bulkquantities}, it is important to realize that $\phi$ comes directly from the NASA Exoplanet Archive parameters, and $F_{XUV}$ comes from the semi-major axis and the stellar SED template. Since the latter is chosen based on the reported stellar spectral type or effective temperature, the XUV flux should be considered an input parameter to our model, and its correlation with $\phi$ simply reveals the underlying observed planet population. The sub-Jovian desert can be vaguely distinguished from it, in the form of a lack of medium-gravity (${\rm log} \; \phi \approx 12.5$ erg g$^{-1}$, i.e., hot Neptune-like) planets at high XUV flux (i.e., close-in). The mass-loss rate is also an input parameter for us, derived from some of the other system parameters, depending most strongly on $F_{XUV}$ and $\phi$. This parametrization of the mass-loss rate is reflected in the correlation with these two parameters: $\dot{M} \propto F_{XUV}$, but at high planet gravity, $\eta_{\rm eff} \propto \dot{M}$ drops dramatically. 

Although the Parker wind temperature is an input parameter to \texttt{sunbather}, it can effectively be considered an output parameter of our model in this case, because we do not a priori know its value and instead solve for it using a self-consistency criterion (see Sect. \ref{sec:model_parameters}). For this reason, it is interesting to see how the self-consistent value depends on the other parameters. By extension, this then also holds for $\lambda$ because it is calculated from $T_0$, $\phi$, and $\bar{\mu}$. From Fig. \ref{fig:correlations_bulkquantities}, we see that $T_0$ depends strongly on the gravitational potential: high-gravity planets develop hot outflows, while low-gravity planets develop relatively cold outflows. This behavior, here seen in the final planet population in a statistical sense, is in line with previous studies on atmospheric escape of individual planets \citep[e.g.,][]{salz_simulating_2016, caldiroli_irradiation-driven_2021, linssen_constraining_2022}. The temperature is naturally also seen to depend on the XUV flux, but the spread is larger than for the planet gravity. The temperature and mass-loss rate are also weakly correlated, where the temperature increases with the mass-loss rate, except at the highest temperatures that are reached at the lowest mass-loss rates. The latter sub-population represents the high-gravity planets, and it may therefore be more intuitive to state that $T_0$ causally depends on $\phi$ and $F_{XUV}$, and is only correlated with $\dot{M}$. However, this distinction is ultimately trivial, because $\dot{M}$ in turn is a function of $\phi$ and $F_{XUV}$. 

In principle, there are many more system parameters that are not shown in Fig. \ref{fig:correlations_bulkquantities}, such as $R_p$, $M_p$, $a$, and $M_*$. However, we found that rather than considering $R_p$ and $M_p$ individually, their combined value in the form of the planet gravity captures the outflow physics just as well. Similarly, $F_{XUV}$ captures the physics better than $a$. $M_*$ directly affects the stellar tidal gravity term, which in most cases only has a minor effect on the outflow physics. $M_*$ also indirectly affects the stellar spectral type and hence the used SED. For the bulk outflow properties that we consider in Fig. \ref{fig:correlations_bulkquantities}, the total XUV flux is more important than the shape of the SED, and is already included. The SED shape has a stronger effect on the microphysics governing the line formation in the planetary transit spectrum, and will be considered in more detail in Sect. \ref{sec:line_correlations}.

\begin{figure}
   \centering
   \includegraphics[width=\hsize]{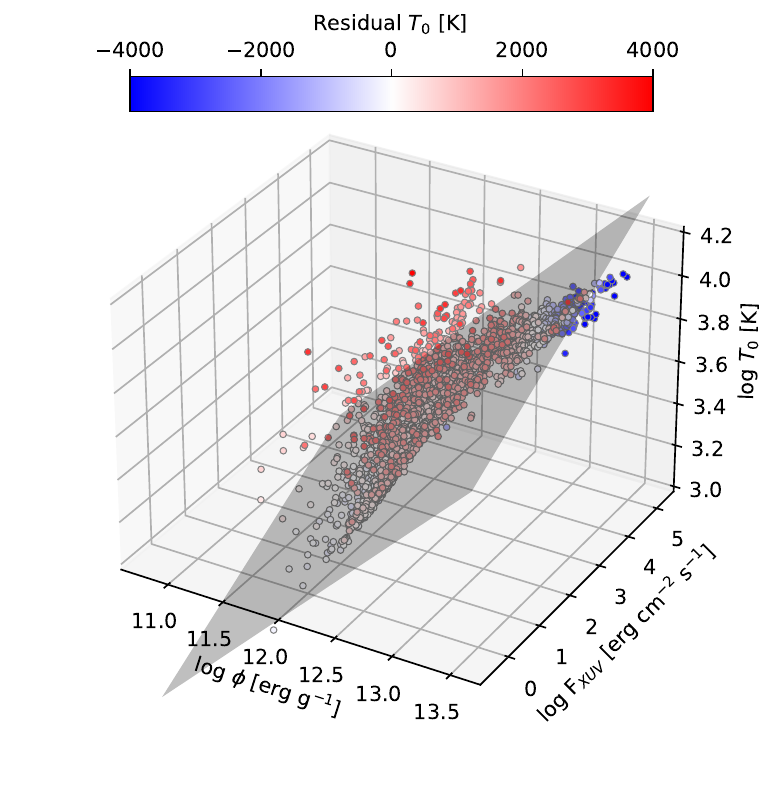}
      \caption{Fit to the isothermal Parker wind temperature, as a function of the planetary gravitational potential and received XUV flux. The final planet population (scatter points) is fitted with a power law (grey plane). The functional form is given by Eq. \ref{eq:T_parametrization}. The systematic residuals reveals that this functional form is not optimal, and particularly does not fully capture the dependence on the gravity.}
         \label{fig:T_fit_3d}
   \end{figure}

While we qualitatively established the dependence of the atmospheric temperature on the planet gravity and the XUV flux before, we find it useful to provide a rough quantitative estimate of this relation. To this end, we fit a simple power law through the final planet population $T_0$-values, as a function of $\phi$ and $F_{XUV}$. This fit is shown in Fig. \ref{fig:T_fit_3d}, and is given by
\begin{equation}
T_0 = 6.64 \times 10^{-4} \; \Bigg( \frac{\phi}{{\rm erg \; g^{-1}}} \Bigg) ^{0.516} \Bigg( \frac{F_{XUV}}{{\rm erg \; cm^{-2} \; s^{-2}}} \Bigg) ^{0.111} \;{\rm K.}
\label{eq:T_parametrization}
\end{equation}
The fit still has considerable structured residuals. It generally underpredicts the temperature of planets with $\phi \approx 12.5$, and overpredicts the temperature of planets with $\phi \gtrsim 13$. The residuals have a mean of 173~K and a standard deviation of 827~K. The fit could not be improved by using a power law dependence on more and/or other system parameters. The functional form of the power law therefore seems to be the reason for the mediocre fit quality: the temperature does not truly depend on the gravitational potential and the XUV flux in this simple way. The power law fit is still useful, because it makes it easy to see how the temperature scales with the two parameters, at least to zeroth order. The $T_0$-parametrization may furthermore find its use in swiftly estimating a prior to break the $T_0-\dot{M}$-degeneracy that is often found in isothermal Parker-wind analyses of spectral line observations \citep[e.g.,][]{vissapragada_constraints_2020, paragas_metastable_2021, allart_homogeneous_2023}. 

\subsection{Spectral line strength correlations} \label{sec:line_correlations}
\begin{figure*}
   \centering
   \includegraphics[width=\hsize]{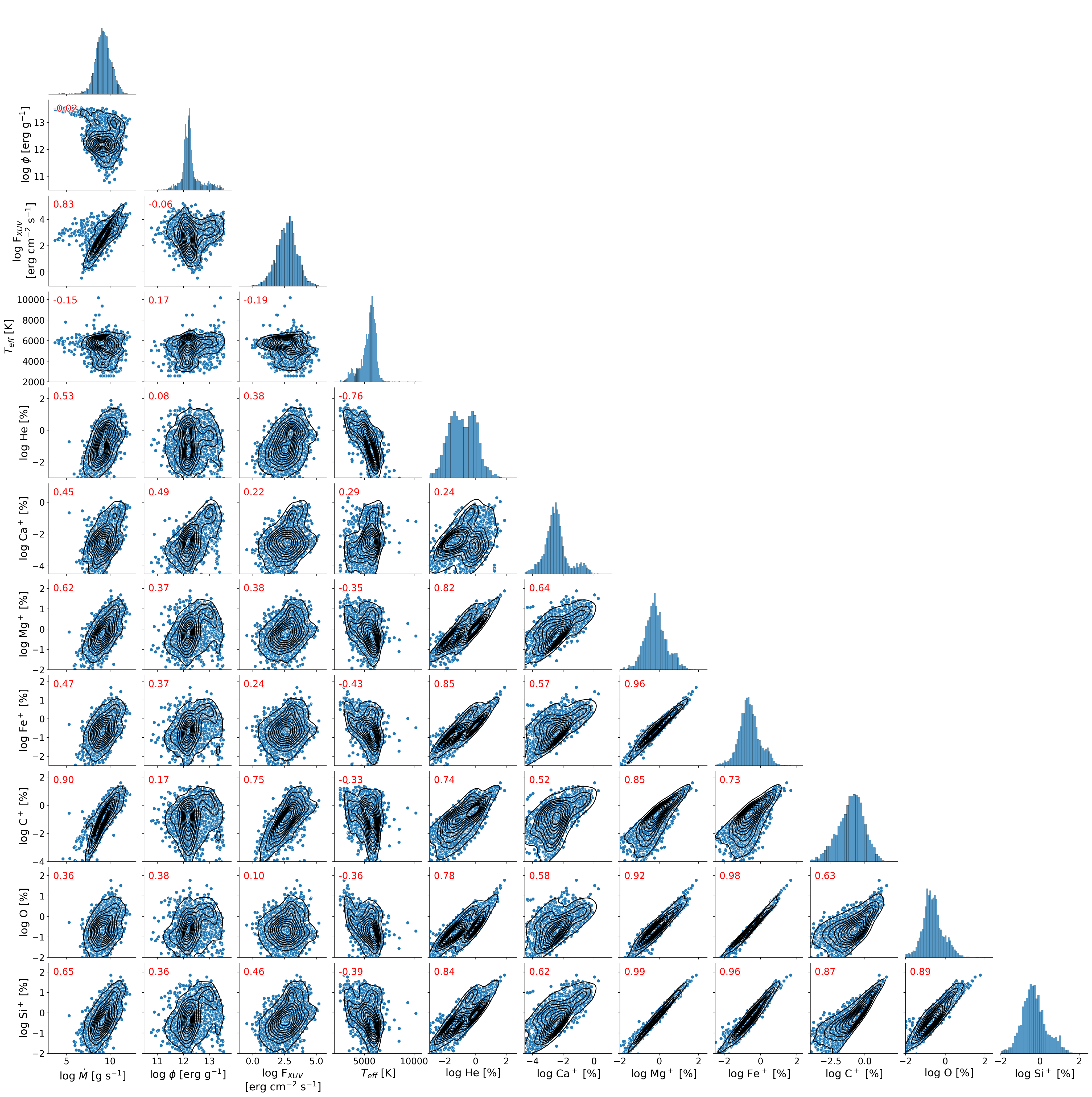}
      \caption{Correlations between various bulk system parameters and spectral line strengths in the transmission spectrum. The spectral line strengths are expressed as the transit depth at line-center. Specifically, they are the metastable helium line evaluated at 10833.25~Å, the calcium infrared triplet evaluated at 8544.44~Å, the magnesium doublet evaluated at 2796.35~Å, the iron UV2 multiplet evaluated at 2382.76~Å, the carbon triplet evaluated at 1335.68~Å, the oxygen triplet evaluated at 1302.17, and the silicon triplet evaluated at 1264.74~Å. Spearman's $\rho$ correlation coefficients are quoted in red in each subplot. They are calculated from the quantities as they are labeled on the axes, so in log-log space except for the stellar effective temperature.}
         \label{fig:correlations_linestrengths}
   \end{figure*}

We now investigate the correlations of various spectral lines. From the transmission spectrum of every planet, we extract the transit depth of the metastable helium triplet (10833~Å), the calcium infrared triplet (8544~Å), the magnesium doublet (2796~Å), the iron UV2 multiplet (2383~Å), the carbon triplet (1336~Å), the oxygen triplet (1302~Å), and the silicon triplet (1265~Å). We show the spectral line correlations of the resulting 305 planets in Fig. \ref{fig:correlations_linestrengths}. We do not analyze the H$\alpha$ line, because the base of the outflow where the transition to hydrostatic equilibrium takes place is not properly modeled in \texttt{sunbather}, while these atmospheric layers are expected to provide substantial absorption in the H$\alpha$ line \citep[e.g.,][]{linssen_expanding_2023}.

The first important realization is that many spectral line strengths show only a moderate correlation with the mass-loss rate. For example, the Spearman's $\rho$ coefficient between the mass-loss rate and helium line strength is 0.53 (in log-log space). This means that the physics governing the line formation are important, and we must be careful not to simply equate helium (non)detections with atmospheric escape (non)detections, without conducting radiative transfer modeling for each system. The helium line strength also does not correlate strongly with the XUV flux ($\rho=0.38$ in log-log space). This goes against the common assumption in the literature. Many works have searched for a trend in the observed exoplanet sample between the received XUV irradiation and the helium line depth (usually normalized by the scale height of the lower atmosphere)  \citep[e.g.,][]{nortmann_ground-based_2018, alonso-floriano_he_2019, kasper_nondetection_2020, zhang_no_2021, zhang_detection_2023, fossati_gaps_2022, kirk_kecknirspec_2022, poppenhaeger_helium_2022, orell-miquel_tentative_2022, orell-miquel_mopys_2024, bennett_nondetection_2023, zhang_giant_2023, guilluy_lv_2024, krishnamurthy_helium_2024}. Whether the correlation can be seen varies in these works, as the number of helium observations has grown over the years, but the most recent versions, based on the largest sample sizes, show no strong correlation. Fig. \ref{fig:correlations_linestrengths} demonstrates that the absence of this correlation is actually in line with the expectations from our models. The helium line strength is however seen to anticorrelate with the stellar effective temperature, which in this case can be considered a proxy for the shape of the SED. Specifically, the EUV/near-UV (NUV) flux ratio varies with stellar spectral type, and is known to influence the metastable helium abundance in the planet atmosphere \citep{oklopcic_helium_2019}.

We move on to fit the helium line equivalent width (EW) as a function of various system parameters. Similar to our fit of the Parker wind temperature, we use a power law dependence, and find that the parameters giving the best fit are the planet mass and radius, the stellar radius, the XUV flux, and the ratio between the ground-state helium ionizing flux ($F_{\rm He-ion}$) and metastable helium ionizing flux ($F_{\rm He^*-ion}$). We ignore the 285 planets with $\phi > 10^{13}$~g~s$^{-1}$, as in that regime, the dependence starts to deviate very strongly from a power law. We obtain
\begin{multline}
    {\rm EW_{He}} = 2.07 \; \Bigg( \frac{R_p}{R_J} \Bigg) ^{3.09} \Bigg( \frac{R_*}{R_{\odot}} \Bigg) ^{-1.57} \Bigg( \frac{M_p}{M_J} \Bigg) ^{-0.864} \\ 
    \cdot \Bigg( \frac{F_{XUV}}{{\rm erg \; cm^{-2} \; s^{-2}}} \Bigg) ^{0.474} \Bigg( \frac{F_{\rm He-ion}}{F_{\rm He^* - ion}} \Bigg) ^{-0.502}  \; {\rm mÅ.}
    \label{eq:He_parametrization}
\end{multline}
We defined $F_{\rm He-ion}$ to be all wavelengths below 504~Å, and $F_{\rm He^*-ion}$ to be between 911 and 2600~Å, but the fit is not sensitive to the chosen lower wavelength boundaries. The fit residuals have a standard deviation of a factor $\sim 2$. In this fit, the helium EW does depend on the XUV flux, despite our previous finding that these quantities are not strongly correlated in the final planet population. This likely means that variation in other system properties prevents a visible and strong correlation with the XUV flux alone.

Next, we phenomenologically identify and describe some trends we observe for certain spectral lines. It is beyond the scope of this paper to provide a microphysical explanation (in terms of the atomic structure, radiative transfer, etc.) for this behavior and we leave this for future work. 

The oxygen line strength does not correlate strongly with any of the system parameters when looking at the Spearman's $\rho$ coefficients, but the figure does show that the strongest oxygen signals are found for planets with intermediate mass-loss rates and late-type host stars. 

The carbon triplet shows the opposite behavior: it correlates extremely strongly with the mass-loss rate. The correlation is weaker -- but still strong -- with the XUV flux, which can influence the carbon line strength indirectly through the mass-loss rate, or directly through ionization of carbon to produce the absorbing C$^+$. The former seems more likely, given the the stronger correlation with the mass-loss rate. This suggests that the microphysics of this line are perhaps not very complex, and that the line simply becomes strong when there is a lot of escaping gas. 

The magnesium, iron, and silicon lines are in an intermediate regime, showing moderate correlation with the mass-loss rate. These lines additionally show modest correlation with the planet gravity. Lastly, the calcium infrared triplet is an interesting case, showing quite weak signals for most planets. However, there is a moderate correlation with the planet gravity, and the shape of this correlation reveals that strong calcium infrared triplet signals are exclusively found around high-gravity planets (and mostly at higher stellar effective temperatures). Therefore, the best chance to observe it seems to be on (ultra-)hot Jupiters orbiting early-type stars \citep[in line with the findings of][]{linssen_expanding_2023}. 

There is also insight to gain from the correlation between two spectral line depths. The lowest correlation is found between the metastable helium line and the calcium infrared triplet. This seems to be the case due to the lines favoring different host star types. The lines thus favor complementary parts of the parameter space of planetary systems. When looking at the correlation between helium and the other spectral lines, we notice that a strong helium signature typically also means that the other spectral lines are strong. However, planets with a weaker helium signature show a larger variety in metal line strengths of multiple orders of magnitude. All of these models consider the same solar composition atmosphere, implying that it is difficult to constrain the escaping atmospheric composition simply from an observed presence and/or absence of these lines: tailored planet models are needed to distinguish the effects of the hydrodynamics from the effects of the composition on the line depth ratios. To a degree, the same reasoning holds for other line combinations that are not highly correlated, such as the carbon and oxygen lines.

In the metal-metal line correlations, we see that some combinations of lines are extremely correlated. Their line depth ratio is apparently insensitive to differences in stellar flux and planetary parameters. The implication is that their line depth ratio is instead an excellent tracer of the atmospheric composition and fractionation. The high correlation between the magnesium and iron lines may particularly be of observational interest, as they can be simultaneously observed using a single HST mode \citep{sing_hubble_2019}. In Appendix \ref{app:good_targets}, we provide a list of promising observational targets for each of the spectral lines discussed.

\section{Comparison to the observed helium sample} \label{sec:comparison_to_sample}
\subsection{Motivation} \label{sec:comparison_motivation}
We now take a closer look at the results for the sample of exoplanets that have been observed in the metastable helium line. We are interested to see if our model predictions are generally consistent with the observational findings. There have been previous works with a similar goal, which typically search for correlations between the observed helium signal strength (or detection/nondetection classification) and some of the physical parameters that are thought to influence atmospheric escape and helium observability, such as the stellar XUV flux (or a proxy thereof) and stellar spectral type. For example, \citet{bennett_nondetection_2023} analyzed a sample of 37 planets, and found a positive correlation between the scale height-normalized helium depth and the stellar $R'_{HK}$ activity indicator. They did not find a correlation with stellar age, metallicity, and rotation rate, nor with the planet's surface gravity and semi-major axis. \citet{krishnamurthy_helium_2024} analyzed the detection status of a sample of 57 planets, and found that most detections occured around K- and G-type stars, and at semi-major axes between 0.03 and 0.08~AU. They also found that few detections occurred for small and low-mass planets, which they interpreted as these mature planets already having been stripped by atmospheric escape. They found no clear trends with stellar age and metallicity. \citet{orell-miquel_mopys_2024} analyzed a sample of 70 planets with He~I~$\lambda$~10833, H$\alpha$, or Lyman-$\alpha$ observations. They found no correlation between the stellar age and the detection status of any of these lines, nor with the helium line equivalent width (EW). They also compared a proxy of the observed mass-loss rate, based on the EW and the stellar radius, to an analytically estimated mass-loss rate, based on the energy-limited prescription with a fixed efficiency (their Fig. 8). They did find a correlation between these proxies when considering the planets with positive helium detections. However, there were also many helium nondetections showing worse agreement with the fit. 

The motivation for us to perform a seemingly similar exercise again is twofold. Firstly, the previous works compared the helium observations only to one or two of the system parameters at a time. While this exercise is certainly useful, it is typically not immediately clear whether the absence of an expected trend means that our theoretical understanding of atmospheric escape is lacking, or whether it can still be explained by variation in other system parameters that dilute the trend. With our approach, we compare directly with planet-specific models that aim to include most of these system parameters at once. This yields a more direct comparison of observation to theory -- at least the physics included in our solar-composition 1D semi-isothermal Parker wind model. Secondly, Fig. \ref{fig:correlations_linestrengths} showed that according to our modeling, some of the presupposed correlations between the helium line strength and system parameters (specifically, the mass-loss rate, the planet gravity, and the XUV irradiation) are actually not very strong in the final planet population. We should thus perhaps not be very surprised that previous works generally did not find such strong correlations. This finding also strengthens the motivation for a direct comparison to planet-specific models.

\subsection{Comparison and interpretation}
\begin{figure*}
   \centering
   \includegraphics[width=\hsize]{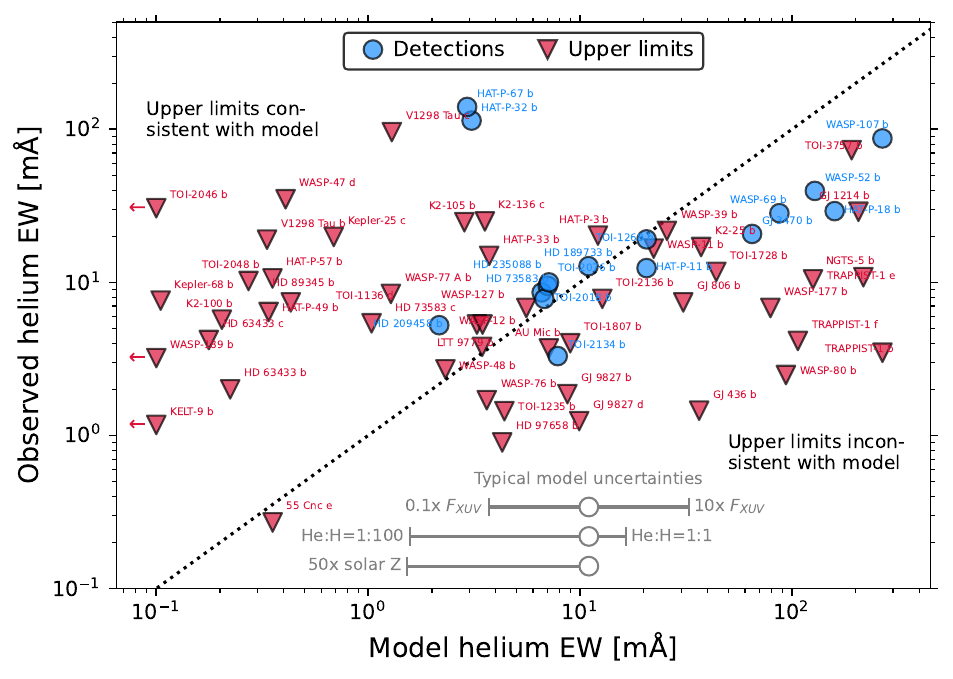}
      \caption{Comparison between the observed and model equivalent width (EW) of the metastable helium line, for the near-complete planet sample as compiled in \citet{orell-miquel_mopys_2024}. The dotted line is $x=y$ (i.e., complete agreement). The model uncertainties are very large. Rough estimates of the uncertainty due to the XUV flux (based on the scaling in Eq. \ref{eq:He_parametrization}), the helium abundance (based on additional simulations of TOI-2018~b), and the metallicity (based on simulations of \citealt{linssen_open-source_2024} of a generic hot Neptune planet) are indicated. An additional significant source of uncertainty is the shape of the stellar SED (specifically the ratio between EUV and NUV flux). Compared to the model uncertainties, observational uncertainties are very small (typically inside the size of the scatter marker) and therefore not shown.}
         \label{fig:mdlHe_vs_obsHe}
   \end{figure*}

We start from the planet sample as compiled in \citet{orell-miquel_mopys_2024}, because it is currently the largest sample in the literature (but may still not be complete). We focus on the 69 planets in the sample with helium observations, and extract the helium line equivalent widths (or upper limits) from their Table M.1. From this sample, we do not have model spectra for four planets: TOI-1683~b is not in the NASA Exoplanet Archive, and our models for KELT-20~b, TOI-1431~b, and K2-77~b failed due to the \texttt{Cloudy} density limit described in Sect. \ref{sec:model_parameters}. For the remaining 65 planets, we integrate the synthetic helium spectrum to obtain the EW, and plot these against the observed values in Fig. \ref{fig:mdlHe_vs_obsHe}. We do not adopt the distinction between non-conclusive and nondetection measurements from \citet{orell-miquel_mopys_2024}, because their designation was based on a comparison to the energy-limited mass-loss rate, which is different from our used mass-loss rate. We caution that the model uncertainties are large, and may also vary significantly from planet to planet, depending on the precision on the input parameters. 

Most of the planets with a positive helium detection have observed EWs relatively close to the model values (within a factor of a few). We consider these consistent, given the large model uncertainties. The exceptions are HAT-P-32~b and HAT-P-67~b, which show almost two orders of magnitude larger absorption than predicted by the model. These planets have small Roche radii of $2.1 R_p$ and $2.6 R_p$, respectively, and are therefore likely in an escape regime where the mass-loss is significantly enhanced over what \texttt{ATES} predicted. The outflow geometry becomes highly non-spherical in this regime, and leading and trailing arms develop. This has indeed been observed for both planets \citep{czesla_h_2022, zhang_giant_2023, bello-arufe_transmission_2023, gully-santiago_large_2024}. These arms and the extra absorption in them are not captured by our model, and could thus lead to the discrepancy. However, some of the other planets in the sample have small Roche radii as well, and using the same reasoning, our model would have underpredicted their absorption too. WASP-67~b, TOI-1807~b, WASP-52~b, HAT-P-57~b, KELT-9~b, LTT~9779~b, WASP-127~b, and WASP-177~b all have Roche radii smaller than $3 R_p$, which would lead to a stronger tension with the observed signals for some of these planets. 

Most of the planets with helium nondetections have upper limits that are above or close to the model predictions, which we consider to be consistent. The planets in the bottom-right part of the figure are those that are most inconsistent and they warrant individual consideration:
\begin{itemize}
    \item The TRAPPIST-1 planets are all likely terrestrial instead of gaseous planets, and are thus not expected to possess extended low mean-molecular weight atmospheres (one of our model assumptions) in the first place. Recent James Webb Space Telescope (JWST) observations have revealed that planets b and c indeed do not possess any substantial atmospheres \citep{greene_thermal_2023, zieba_no_2023}, and pre-JWST results also did not detect any atmospheric features in any of the seven planets \citep[see references in][]{greene_thermal_2023}. If these planets did possess thicker atmospheres in the past, they have likely been eroded already by atmospheric escape \citep{hori_trappist-1_2020, gialluca_implications_2024}.
    \item WASP-80~b is a hot Jupiter planet, and one of the most curious nondetections in the sample. A dedicated work aiming to explain the nondetection concluded that it likely resulted from a low stellar EUV flux, or alternatively, a low helium abundance \citep{fossati_possible_2023}.
    \item GJ~436~b is another peculiar case, as it has been observed to host an extensive escaping atmosphere in the hydrogen Lyman-$\alpha$ line \citep{kulow_ly_2014, ehrenreich_giant_2015, lavie_long_2017}. \citet{rumenskikh_mysterious_2023} modeled the planet in detail and found that the helium nondetection was likely due to the small planet-to-star radius ratio, the radiation pressure folding the outflow along the line-of-sight so that it absorbs over a smaller area, and a helium abundance smaller than 0.3 times the solar value. The small planet-to-star ratio is naturally included in our model, leaving us with the latter two possible explanations for the discrepancy.
    \item The NGTS-5~b observations are from \citet{vissapragada_upper_2022}, who reported a tentative detection at the $2.2 \sigma$ level, which was classified as a nondetection by \citet{orell-miquel_mopys_2024}. This means that the observed EW should be thought of as a measured value rather than an upper limit. The NASA Exoplanet Archive does not report a transit impact parameter for this planet, and our modeling framework thus assumed 0. \citet{vissapragada_upper_2022} reported a value of $b \approx 0.68$, which would have resulted in a lower model EW. However, this difference is only minor and likely less important than other model uncertainties such as the EUV flux and the helium abundance. We do not know specifically which of the model uncertainties could offer a satisfying explanation. We do note that the inconsistency could also be partly due to the data quality. NGTS-5 is a relatively faint star for helium studies with a J-magnitude of 12.1, and \citet{vissapragada_upper_2022} reported difficulties with the data reduction for the first of their two nights of observations. Additional observations with a large telescope would prove very useful in obtaining more precise observational constraints.
    \item WASP-177~b is a highly inflated hot Jupiter with large radius uncertainties of $R_p = 1.58^{+0.66}_{-0.36} R_J$. For a smaller planet radius, the model EW decreases and would be more in line with the observations. Another more exotic possible explanation could be the 3D geometry of the escaping gas. WASP-177~b is on a grazing orbit with $b=0.98$, so that if the stellar wind were to shape the planetary outflow into a cometary tail oriented radially away from the star, a large portion of the escaping gas would not obscure the stellar disk. This scenario would not be compatible with a redshifted helium line, however, which \citet{kirk_kecknirspec_2022} saw tentative evidence of. 
\end{itemize}

We also shortly discuss the planets in the sample that we failed to make model spectra for. KELT-20~b and TOI-1431~b are high-gravity super-Jupiter planets, and their mass-loss rates from the \texttt{ATES} formula are only $\mathcal{O}$($10^4$~g~s$^{-1}$). Such low mass-rates would not produce observable helium signatures, consistent with their observed nondetections. K2-77~b is a $0.205 R_J$-sized planet, but does not have a good mass measurement. The NASA Exoplanet Archive currently quotes a rather unphysical value of $1.9 M_J$. It is unclear if a more reliable mass measure would produce a model helium spectrum that is consistent with the observed nondetection upper limit. 

\begin{figure}
   \centering
   \includegraphics[width=\hsize]{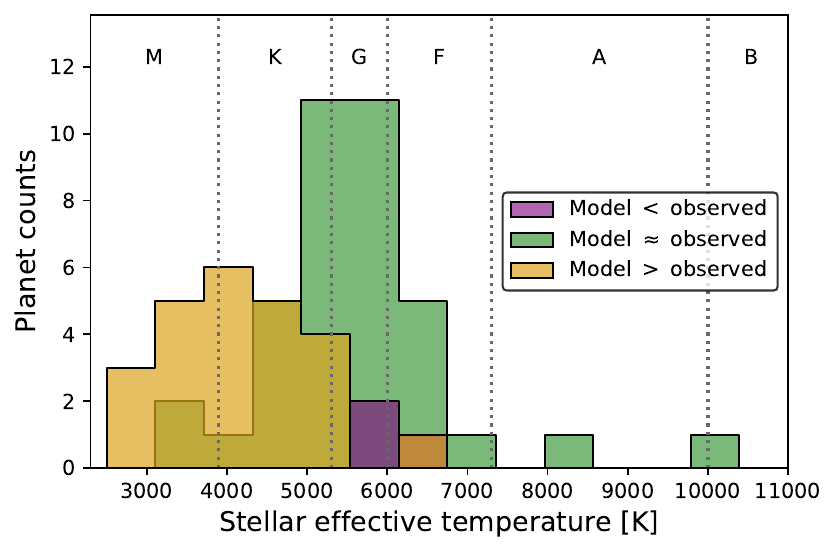}
      \caption{Host star effective temperatures of the planets with metastable helium observations of Fig. \ref{fig:mdlHe_vs_obsHe}, separated by the accuracy of the model predictions. 
      In this figure, we adopt a relatively strict factor of two when classifying a model as accurate. So, the yellow histogram includes the planets for which the model helium EW is more than a factor two higher than the observed EW (i.e., both detections and upper limits towards the bottom-right part of Fig. \ref{fig:mdlHe_vs_obsHe}). The green histogram includes the planets that are accurately predicted by our model: detections with a model EW within a factor two of the observed value, as well as observed upper limits above half the model value (i.e., upper limits towards the top-left part of Fig. \ref{fig:mdlHe_vs_obsHe}). The purple histogram includes the planets with detections that are above two times the model EW (i.e., HAT-P-67~b, HAT-P-32~b, and HD~209458~b). We see that our model tends to overestimate the helium signal particularly for planets around M- and K-type stars.}
         \label{fig:mdlHe_prediction_accuracy}
   \end{figure}

We notice that at high model helium EW, there appears to be a small systematic offset for the planets with confirmed helium detections: our model overpredicts the signal strength of WASP-107~b, HAT-P-18~b, WASP-52~b, WASP-69~b, and GJ~3470~b by a factor of a few. This is still within the expected model uncertainties, so we consider these model predictions to be consistent with the observations, but the systematic nature of the offsets does raise interest. Although speculative, the cause of this behavior potentially lies in the stellar SEDs. The host star spectral types of all planets for which our model overpredicts the helium EW by more than a factor 2 (including both detections and nondetections) are mainly M- and K-type (Fig. \ref{fig:mdlHe_prediction_accuracy}). The remaining planets in the sample which are not overpredicted orbit mainly late-K and earlier type stars. Therefore, it could be the case that the SED templates that were used for the K- and M-type stars are less representative of the true SED of the host stars, although we do not know in which way specifically. 

Most observed helium spectra are consistent with the model predictions -- at least within the large model uncertainties -- and we have decent explanations for the ones that are strongly inconsistent. This leads us to suggest that the general theoretical and modeling framework can sufficiently explain the current metastable helium census. That is, we do not see strong evidence that the observed planets are systematically poorly described by a 1D hydrodynamic photoevaporation model. Of course, our comparison is only based on the equivalent width, and spectrally resolved line shapes do in many cases show evidence of additional physics shaping the planetary outflow, which can only be studied with 3D simulations \citep[e.g.][]{wang_metastable_2021, macleod_stellar_2022, nail_effects_2024}. Still, according to our first-order comparison, photoevaporation is able to explain the present-day metastable helium census.

\subsection{Factoring in observability: the ``helium TSM''}
A slightly different way to compare to the observed helium sample, is to modulate the model predictions by a proxy for the signal-to-noise ratio (S/N) of observations. The S/N depends on many factors, including the magnitude or distance of the system, the telescope used, the weather, and the data reduction algorithms. Out of these, the magnitude of the host star is the only factor that will be constant in each observation. Therefore, we set up a metric for the expected helium observability, based on the model equivalent width of the helium line and the host star's J-magnitude. We use the J-magnitude because it is relatively close to the metastable helium line, and is well-constrained for all planets by the Transiting Exoplanet Survey Satellite \citep{stassun_revised_2019}. We use a similar dependence on the magnitude as the transmission spectroscopy metric (TSM) for the lower atmosphere as defined in \citet{kempton_framework_2018}. Our ``helium TSM'' is defined as
\begin{equation} \label{eq:helium_TSM}
    {\rm helium \; TSM} = \frac{{\rm EW_{He}}}{10833.25 \; \AA} \times 10^{-m_J / 5},
\end{equation}
where ${\rm EW_{He}}$ is the equivalent width of the helium triplet and $m_J$ is the J-magnitude. We divide by the central wavelength of the helium line in order to obtain a normalized and dimensionless TSM metric.

\begin{figure*}
   \centering
   \includegraphics[width=\hsize]{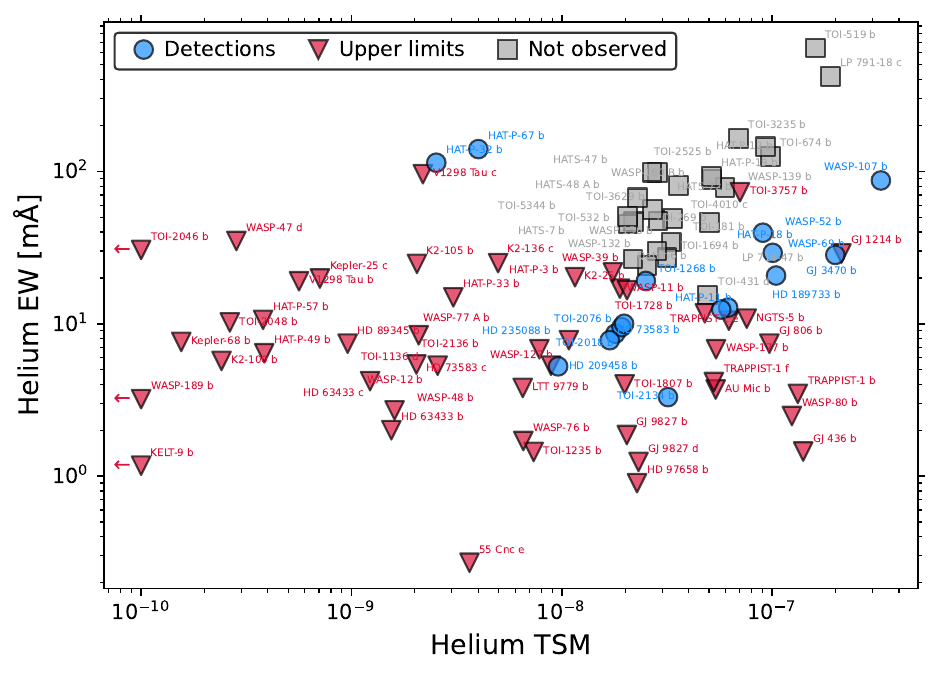}
      \caption{Comparison between the equivalent width (EW) of the metastable helium line and the helium TSM (a metric for the expected helium observability based on the model EW and the J-magnitude; Eq. \ref{eq:helium_TSM}). We include the observed helium planet sample as compiled in \citet{orell-miquel_mopys_2024}, as well as potentially good targets that have not yet been observed and have helium TSM $> 2\times10^{-8}$. For the observed planets, we plot observed EW values, while we plot model EW values for the planets that have not been targeted yet. The relatively high EW values (but comparable helium TSM values) of the potentially good targets compared to the observed sample means that the former are generally orbiting fainter stars. They are also listed in Table \ref{tab:best_He_TSM}.}
         \label{fig:HeTSM_vs_obsEW}
   \end{figure*}

In Fig. \ref{fig:HeTSM_vs_obsEW}, we present the helium TSM values for the planets in the sample of \citet{orell-miquel_mopys_2024}. The figure looks qualitatively similar to Fig. \ref{fig:mdlHe_vs_obsHe}, but the detected and nondetected planet subsamples separate out slightly more distinctly when plotted against the helium TSM as done in Fig. \ref{fig:HeTSM_vs_obsEW}. We interpret this as another indication that the general theoretical framework of photoevaporation seems to have predictive power over metastable helium line observations. Most helium detections thus far have been for planets with a helium TSM $\gtrsim 2\times10^{-8}$. In our final population, there are many planets with a helium TSM above this threshold, which are potentially good observational targets. If we filter out the planets that have serious model uncertainties or warnings (these are listed in Appendix \ref{app:good_targets}), we are left with 25 unobserved planets with a helium TSM above $2\times10^{-8}$. These are also plotted in Fig. \ref{fig:HeTSM_vs_obsEW}. 

We do not compare our model spectra in different spectral lines to their respective observed planet samples. The Lyman-$\alpha$ line is not reliably predicted by our spherically symmetric model, as it is strongly shaped by interactions with the stellar wind and radiation pressure. The H$\alpha$ line depth is systematically underpredicted by our model, because it has a significant contribution from the quasi-hydrostatic lower atmospheric layers that are not properly modeled in the isothermal Parker wind profile \citep[see also][]{linssen_expanding_2023}. The sample size of observations of metal lines is too small to do a meaningful model comparison. However, we do define specific TSM metrics for the calcium infrared triplet, the magnesium NUV doublet, and the carbon and oxygen FUV triplets, and list some of the most promising candidates in Appendix \ref{app:good_targets}.

\section{Summary} \label{sec:summary}
We perform numerical modeling with the \texttt{sunbather} code, to produce synthetic transmission spectra of the upper atmospheres of most transiting exoplanets. The starting point is the NASA Exoplanet Archive, from which we select every transiting planet with reported values for the necessary system parameters. We use a set of stellar SED templates that is mainly comprised of SEDs from the MUSCLES survey. For each planet, the assumed atmospheric mass-loss rate comes from the analytic approximation of \citet{caldiroli_irradiation-driven_2022}, which is based on modeling with the photoevaporation code \texttt{ATES}. The isothermal Parker wind temperature is chosen based on a self-consistency argument. We assume a solar composition atmosphere. Using all of these ingredients, we calculate high-resolution multi-wavelength (911 to 11,000~Å) transit spectra of 3898 exoplanets (246 models fail). These spectra are hosted online in the \texttt{sunset} database$^{\ref{fn:data}}$.

The large number of transit spectra allows us to study how the strength of spectral lines relates to the bulk planetary parameters. The metastable helium triplet is found to correlate only moderately with the mass-loss rate, implying that we cannot always equate a nondetection to the absence of atmospheric escape -- even for a solar helium abundance. We also find the helium line strength to correlate only weakly with the stellar XUV flux, suggesting that the absence of such a trend in observational studies may, in fact, be expected. Other spectral lines such as the calcium infrared triplet and iron NUV multiplet show similar moderate correlations with the mass-loss rate. The exception is the O~I~1302~Å triplet, which is found to correlate strongly with the mass-loss rate. Line-line correlations reveal that the calcium infrared triplet prefers a part of the planetary parameter space (high-gravity planets around early-type stars) that is highly complementary to that of the other lines. Of particular observational interest is the strong correlation between the magnesium doublet and iron multiplet (as they can be targeted in a single HST mode), indicating that their line ratio is an excellent probe of their abundance ratio.

We also compare our synthetic spectra to a near-complete sample of 65 observed metastable helium spectra. Save a few interesting exceptions, our results are generally consistent with the observed census, within the large model uncertainties. This suggests that, to first degree, photoevaporation can explain the present-day landscape of metastable helium observations. Finally, for each of the spectral lines focused on in this work, we define a transmission spectroscopy metric based on the line strength and system magnitude or distance. These metrics can help to identify the best observational targets.

\begin{acknowledgements}
      We thank L. dos Santos, J. Spake, and F. Nail for the helpful conversations. This work has made use of the Snellius Supercomputer operated by SURF. A. Oklop\v{c}i\'{c} gratefully acknowledges support from the Dutch Research Council NWO Veni grant.
\end{acknowledgements}

\bibliographystyle{aa}
\bibliography{library}

\begin{appendix}
\section{Promising targets for various spectral lines} \label{app:good_targets}
Here, we list planets that our model predicts to be favorable targets for different spectral lines. For this, we only include the 326 exoplanets without the potentially serious warnings listed at the end of Sec. \ref{sec:warnings}. For the metastable helium line, we rank these planets by their helium TSM (Eq. \ref{eq:helium_TSM}). In table \ref{tab:best_He_TSM}, we list all planets with helium TSM $> 2\times10^{-8}$. WASP-107~b, which was seen to have the highest helium TSM in Fig. \ref{fig:HeTSM_vs_obsEW}, is not in the table because it has the warning that its $K \phi$ value is below the validity bounds of the \citet{caldiroli_irradiation-driven_2022} formula. For the calcium infrared triplet, we define a similar metric:
\begin{equation} \label{eq:calcium_TSM}
    {\rm calcium \; TSM} = \frac{{\rm EW_{Ca}}}{8544.44 \; \AA} \times 10^{-m_V / 5},
\end{equation}
where ${\rm EW_{Ca}}$ is the equivalent in the calcium line, and $m_V$ is the V-magnitude (which is available for each planet and closer to the line than the J-magnitude). In table \ref{tab:best_Ca_TSM}, we list the ten planets with the highest calcium TSM. For the magnesium NUV doublet, the carbon FUV triplet, and the oxygen FUV triplet, we also define TSM metrics. The shortest wavelength photometry that is available for most systems is the B-magnitude, which is likely not a good proxy for the S/N at NUV and FUV wavelengths. Therefore, we simply use the system distance as a proxy for the S/N, and define the respective TSM metrics as
\begin{equation} \label{eq:UVline_TSM}
    {\rm magnesium/carbon/oxygen \; TSM} = \frac{{\rm EW_{line}}}{\lambda_{\rm line}} \times \frac{7.7 \; {\rm pc}}{d},
\end{equation}
where ${\rm EW_{line}}$ and $\lambda_{\rm line}$ are the equivalent width and central wavelength of the line, and $d$ is the distance to the system. We include a factor of 7.7~pc to normalize to the distance of the star Vega, as this star is also used to normalize the other apparent magnitudes. We list the ten planets with the highest TSM in each line in table \ref{tab:best_Mg_TSM}, \ref{tab:best_C_TSM}, and \ref{tab:best_O_TSM}. We point out that the different definitions of the helium, calcium, and UV line TSM metrics means they cannot be compared quantitatively. These metrics are instead intended to be used independently to compare different planets within each metric.

\begin{table}
\caption{Best targets for the metastable helium triplet at 10833~Å. The transmission spectroscopy metric is defined in Eq. \ref{eq:helium_TSM}. WASP-107~b and a few other planets that have a high helium TSM in Fig. \ref{fig:HeTSM_vs_obsEW}, are not in this table because they have potentially serious warnings described at the end of Sec. \ref{sec:warnings}. See \citet{orell-miquel_mopys_2024} for references on the observed planets. TOI~1259~A~b is not in their sample, but was observed by \citet{saidel_atmospheric_2024}. The helium TSM values of all planets can be found in the \texttt{sunset} database$^{\ref{fn:data}}$.}
\label{tab:best_He_TSM}
\begin{tabular}{l l l}
\hline\hline
Planet & Helium TSM & Observations published \\
\hline
WASP-69 b & $1.99 \times 10^{-7}$ & Yes\\
LP 791-18 c & $1.89 \times 10^{-7}$ & No\\
TOI-519 b & $1.60 \times 10^{-7}$ & No\\
GJ 436 b & $1.40 \times 10^{-7}$ & Yes\\
WASP-80 b & $1.24 \times 10^{-7}$ & Yes\\
HAT-P-18 b & $1.01 \times 10^{-7}$ & Yes\\
TOI-674 b & $9.82 \times 10^{-8}$ & No\\
HAT-P-12 b & $9.28 \times 10^{-8}$ & No\\
WASP-52 b & $9.03 \times 10^{-8}$ & Yes\\
NGTS-5 b & $7.56 \times 10^{-8}$ & Yes\\
TOI-3757 b & $7.02 \times 10^{-8}$ & Yes\\
TOI-1259 A b & $6.93 \times 10^{-8}$ & Yes\\
TOI-3235 b & $6.92 \times 10^{-8}$ & No\\
HD 189733 b & $6.17 \times 10^{-8}$ & Yes\\
HATS-72 b & $5.97 \times 10^{-8}$ & No\\
HAT-P-11 b & $5.72 \times 10^{-8}$ & Yes\\
AU Mic b & $5.38 \times 10^{-8}$ & Yes\\
WASP-139 b & $5.18 \times 10^{-8}$ & No\\
HAT-P-19 b & $5.16 \times 10^{-8}$ & No\\
TOI-181 b & $5.04 \times 10^{-8}$ & No\\
TOI-431 d & $4.94 \times 10^{-8}$ & No\\
WASP-160 B b & $3.58 \times 10^{-8}$ & No\\
TOI-4010 c & $3.37 \times 10^{-8}$ & No\\
WASP-156 b & $3.31 \times 10^{-8}$ & No\\
TOI-2134 b & $3.20 \times 10^{-8}$ & Yes\\
LP 714-47 b & $3.16 \times 10^{-8}$ & No\\
TOI-269 b & $2.87 \times 10^{-8}$ & No\\
TOI-2525 b & $2.84 \times 10^{-8}$ & No\\
TOI-1694 b & $2.83 \times 10^{-8}$ & No\\
TOI-3629 b & $2.69 \times 10^{-8}$ & No\\
HATS-47 b & $2.69 \times 10^{-8}$ & No\\
Qatar-6 b & $2.54 \times 10^{-8}$ & No\\
TOI-1268 b & $2.51 \times 10^{-8}$ & Yes\\
HATS-48 A b & $2.29 \times 10^{-8}$ & No\\
TOI-532 b & $2.19 \times 10^{-8}$ & No\\
WASP-132 b & $2.18 \times 10^{-8}$ & No\\
HATS-7 b & $2.07 \times 10^{-8}$ & No\\
TOI-5344 b & $2.05 \times 10^{-8}$ & No\\
\hline
\end{tabular}
\end{table}

\begin{table}
\caption{Best targets for the calcium infrared triplet at 8544~Å. The transmission spectroscopy metric is defined in Eq. \ref{eq:calcium_TSM}. The calcium TSM values of all planets can be found in the \texttt{sunset} database$^{\ref{fn:data}}$.}
\label{tab:best_Ca_TSM}
\begin{tabular}{l l l}
\hline\hline
Planet & Calcium TSM & Observations published \\
\hline
WASP-33 b & $1.47 \times 10^{-9}$ & \citet{yan_ionized_2019} \\
HD 209458 b & $1.21 \times 10^{-9}$ & No\\
KELT-7 b & $1.17 \times 10^{-9}$ & No\\
WASP-121 b & $9.10 \times 10^{-10}$ & No (but Ca~II~H\&K by\\
& & \citealt{borsa_atmospheric_2021})\\
KELT-4 A b & $8.84 \times 10^{-10}$ & No\\
HD 189733 b & $7.68 \times 10^{-10}$ & No\\
WASP-79 b & $7.46 \times 10^{-10}$ & No\\
WASP-101 b & $7.17 \times 10^{-10}$ & No\\
KELT-23 A b & $6.21 \times 10^{-10}$ & No\\
WASP-7 b & $6.01 \times 10^{-10}$ & No\\
\hline
\end{tabular}
\end{table}

\begin{table}
\caption{Best targets for the magnesium doublet at 2796~Å, and likely also for the UV2 iron multiplet and the silicon line at 1265~Å, since these line strengths are highly correlated. The transmission spectroscopy metric is defined in Eq. \ref{eq:UVline_TSM}. The magnesium TSM values of all planets can be found in the \texttt{sunset} database$^{\ref{fn:data}}$.}
\label{tab:best_Mg_TSM}
\begin{tabular}{l l l}
\hline\hline
Planet & Magnesium TSM & Observations published \\
\hline
LP 791-18 c & $6.72 \times 10^{-6}$ & No\\
TOI-519 b & $3.10 \times 10^{-6}$ & No\\
GJ 436 b & $2.14 \times 10^{-6}$ & No\\
TOI-674 b & $1.70 \times 10^{-6}$ & No\\
WASP-69 b & $1.61 \times 10^{-6}$ & No\\
WASP-80 b & $1.30 \times 10^{-6}$ & No\\
TOI-3235 b & $1.14 \times 10^{-6}$ & No\\
HAT-P-12 b & $9.79 \times 10^{-7}$ & No\\
HAT-P-18 b & $8.35 \times 10^{-7}$ & No\\
WASP-52 b & $7.71 \times 10^{-7}$ & No\\
\hline
\end{tabular}
\end{table}

\begin{table}
\caption{Best targets for the carbon triplet at 1336~Å. The transmission spectroscopy metric is defined in Eq. \ref{eq:UVline_TSM}. The carbon TSM values of all planets can be found in the \texttt{sunset} database$^{\ref{fn:data}}$.}
\label{tab:best_C_TSM}
\begin{tabular}{l l l}
\hline\hline
Planet & Carbon TSM & Observations published \\
\hline
LP 791-18 c & $6.19 \times 10^{-6}$ & No\\
TOI-519 b & $3.23 \times 10^{-6}$ & No\\
GJ 436 b & $1.60 \times 10^{-6}$ & \citet{loyd_ultraviolet_2017};\\
& & \citet{dos_santos_hubble_2019}\\
TOI-3235 b & $1.48 \times 10^{-6}$ & No\\
TOI-674 b & $1.43 \times 10^{-6}$ & No\\
WASP-80 b & $1.43 \times 10^{-6}$ & No\\
WASP-69 b & $1.24 \times 10^{-6}$ & No\\
WASP-52 b & $8.18 \times 10^{-7}$ & No\\
HAT-P-12 b & $7.38 \times 10^{-7}$ & No\\
HD 189733 b & $7.15 \times 10^{-7}$ & \citet{ben-jaffel_hubble_2013}\\
\hline
\end{tabular}
\end{table}

\begin{table}
\caption{Best targets for the oxygen triplet at 1302~Å. The transmission spectroscopy metric is defined in Eq. \ref{eq:UVline_TSM}. The oxygen TSM values of all planets can be found in the \texttt{sunset} database$^{\ref{fn:data}}$.}
\label{tab:best_O_TSM}
\begin{tabular}{l l l}
\hline\hline
Planet & Oxygen TSM & Observations published \\
\hline
LP 791-18 c & $4.94 \times 10^{-6}$ & No\\
TOI-519 b & $1.40 \times 10^{-6}$ & No\\
TOI-3235 b & $7.17 \times 10^{-7}$ & No\\
GJ 436 b & $7.07 \times 10^{-7}$ & No\\
TOI-674 b & $4.77 \times 10^{-7}$ & No\\
WASP-80 b & $1.93 \times 10^{-7}$ & No\\
HD 189733 b & $1.66 \times 10^{-7}$ & \citet{ben-jaffel_hubble_2013} \\
WASP-69 b & $1.61 \times 10^{-7}$ & No\\
TOI-269 b & $1.41 \times 10^{-7}$ & No\\
TOI-3757 b & $1.31 \times 10^{-7}$ & No\\
\hline
\end{tabular}
\end{table}  

\end{appendix}

\end{document}